\documentclass[onecolumn,12pt]{article}
\usepackage{graphicx}
\newcommand\simlt{\hspace{0.3em}\raisebox{0.4ex}{$<$}\hspace{-0.75em}\raisebox{-.7ex}{$\sim$}\hspace{0.3em}}

\begin{document}
\baselineskip 14pt

\title{Distortion of Magnetic Fields in a Starless Core IV: \\
Magnetic Field Scaling on Density and \\
Mass-to-flux Ratio Distribution in FeSt 1-457} 
\date{}
\author{Ryo Kandori$^{1}$, Kohji Tomisaka$^{2}$, Motohide Tamura$^{1,2,3}$, Masao Saito$^{2}$, \\
Nobuhiko Kusakabe$^{1}$, Yasushi Nakajima$^{4}$, Jungmi Kwon$^{5}$, Takahiro Nagayama$^{6}$, \\
Tetsuya Nagata$^{7}$, and Ken'ichi Tatematsu$^{2}$\\
{\small 1. Astrobiology Center of NINS, 2-21-1, Osawa, Mitaka, Tokyo 181-8588, Japan}\\
{\small 2. National Astronomical Observatory of Japan, 2-21-1 Osawa, Mitaka, Tokyo 181-8588, Japan}\\
{\small 3. Department of Astronomy, The University of Tokyo, 7-3-1, Hongo, Bunkyo-ku, Tokyo, 113-0033, Japan}\\
{\small 4. Hitotsubashi University, 2-1 Naka, Kunitachi, Tokyo 186-8601, Japan}\\
{\small 5. Institute of Space and Astronautical Science, Japan Aerospace Exploration Agency,}\\
{\small 3-1-1 Yoshinodai, Chuo-ku, Sagamihara, Kanagawa 252-5210, Japan}\\
{\small 6. Kagoshima University, 1-21-35 Korimoto, Kagoshima 890-0065, Japan}\\
{\small 7. Kyoto University, Kitashirakawa-Oiwake-cho, Sakyo-ku, Kyoto 606-8502, Japan}\\
{\small e-mail: r.kandori@nao.ac.jp}}
\maketitle

\abstract{\bf 
In the present study, the magnetic field scaling on density, $|B| \propto \rho^{\kappa}$, was revealed in a single starless core for the first time. The $\kappa$ index of $0.78 \pm 0.10$ was obtained toward the starless dense core FeSt 1-457 based on the analysis of the radial distribution of the polarization angle dispersion of background stars measured at the near-infrared wavelengths. The result prefers $\kappa = 2/3$ for the case of isotropic contraction, and the difference of the observed value from $\kappa = 1/2$ is 2.8 sigma. The distribution of the ratio of mass to magnetic flux was evaluated. FeSt 1-457 was found to be magnetically supercritical near the center ($\lambda \approx 2$), whereas nearly critical or slightly subcritical at the core boundary ($\lambda \approx 0.98$). Ambipolar-diffusion-regulated star formation models for the case of moderate magnetic field strength may explain the physical status of FeSt 1-457. The mass-to-flux ratio distribution for typical dense cores (critical Bonnor--Ebert sphere with central $\lambda=2$ and $\kappa=1/2$--$2/3$) was calculated and found to be magnetically critical/subcritical at the core edge, which indicates that typical dense cores are embedded in and evolve from magnetically critical/subcritical diffuse surrounding medium. 
}

\vspace*{0.3 cm}

\clearpage

\section{Introduction}
Magnetic fields are believed to play an important role in controlling the formation and contraction of dense cores in molecular clouds. The determination of the relationships between the magnetic field strength $|B|$ and the gas volume density $\rho$, usually expressed in a power law form as $|B| \propto \rho^{\kappa}$, is important because they are related to the accumulation history of both the magnetic flux and the cloud material (e.g., Crutcher 1999). The $|B|$--$\rho$ relationship is also crucial in order to compare the magnetic field and internal density structure observations with theory. 
\par
If an initially uniform magnetic field pervading a diffuse medium is assumed as a starting condition of the mass accumulation to form dense cores, the $|B|$--$\rho$ relationship of the core depends on 1) the shape of the progenitor cloud (e.g., spherical, cylindrical), 2) the magnetic field geometry (i.e., parallel or perpendicular or inclined geometry with respect to the elongation axis of the core), and 3) the direction of contraction (i.e., isotropic contraction or contraction toward a specific direction). 
In the case of (spherical) isotropic contraction, the conservation of magnetic flux ($\Phi = \pi R^2 |B|$) yields $|B| \propto R^{-2}$ ($R$ is the radius of the core) and the conservation of mass $(M = (4/3)\pi R^3 \rho)$ yields $\rho^{2/3} \propto R^{-2}$, providing the $|B|$--$\rho$ relationship as $|B| \propto \rho^{2/3}$. This corresponds to the prediction of the relatively weak magnetic field case (Mestel 1966). Note that isotropic contraction does not necessarily mean spherical cloud shape, merely that the shape be conserved during the contraction. However, if the initial axial ratio of the cloud is large, the shape of the cloud becomes more elongated during the contraction by the effect of gravity. 
%
%
In the case of the plane-parallel or infinite thin disk geometry, the conservation of magnetic flux ($\Phi = \pi R^2 |B|$) and mass $(M = \pi R^2 z \rho)$ yields $\rho z / |B| = {\rm constant}$, where $z$ is the distance perpendicular to the plane. In this geometry, the force balance between self-gravity and internal thermal pressure along the symmetry axis is $2\pi G \rho z^2 \approx C_{\rm s}^2$ (Spitzer 1942), where $C_{\rm s}$ is the sound speed. Therefore, $|B| \propto (\rho T)^{1/2}$ ($T$ is the gas temperature), and in the isothermal case, $|B| \propto \rho^{1/2}$ (see, Crutcher 1999). 
\par 
On the basis of large samples with Zeeman measurements of the line-of-sight magnetic field strength $B_{\rm los}$ and Bayesian statistical analysis, Crutcher et al. (2010) concluded that the data prefer $\kappa \approx 2/3$ $(|B| \propto \rho^{0.65 \pm 0.05}$ for $\rho > 300$ cm$^{-3}$) and reject $\kappa \approx 1/2$. They also showed the existence of two distinct branches on the $B$ versus $\rho$ diagram, a flat region at low densities ($|B|$ independent of $\rho$, i.e., $\kappa \approx 0$) and a power-law scaling region at high densities ($\kappa \approx 2/3$). 
A recent study reported results contrary to those reported by Crutcher et al. (2010) based on the re-analysis of the same observational data ($\kappa \approx 1/2$ is preferred; Tritsis et al. 2015). Note that Crutcher et al. (2010) analyzed the full set of Zeeman data including non-detections, whereas Tritsis et al. (2015) only analyzed the observational data with Zeeman detection (this may cause the biased results with stronger magnetic field strength and smaller $\kappa$). 
\par
Several $\kappa$ measurements with smaller samples have been conducted. Li et al. (2015) obtained $\kappa = 0.41 \pm 0.04$ toward the clouds and cores in the NGC 6334 complex based on the measurements of $B_{\rm pos}$ by comparing the curvature of the plane-of-sky magnetic field lines with self-gravity. Ching et al. (2017) obtained $\kappa = 0.54 \pm 0.30$ toward the cores in the dense filamentary cloud DR21 based on the submillimeter (submm) dust emission polarimetry and the Chandrasekhar--Fermi method (Chandrasekhar \& Fermi 1953). Hoq et al. (2017) obtained $\kappa = 0.73 \pm 0.06$ toward the filamentary infrared dark cloud (IRDC) G28.23-00.19 based on near-infrared (NIR) dust extinction polarimetry and the Chandrasekhar--Fermi method. 
Observations show a variety of $\kappa$ values ranging from $\kappa \approx 1/2$ to $\kappa \approx 2/3$. Therefore, it is important for observational studies to provide the definite value of $\kappa$ through much larger samples or much more accurate measurements, although it is possible that the value of $\kappa$ varies from region to region, depending on the shape of objects or the type of contractions or other characteristics. Note that there is no observation of the $|B|$--$\rho$ relationship determined using a single molecular cloud core. 
\par
From a theoretical point of view, Mouschovias (1976a,b) showed that the ratio of magnetic and gas pressure ($B^2 / 8 \pi P$) tends to remain constant, $\approx 1$, inside the magnetized cloud during collapse. This yields $|B| \propto \rho^{1/2}$ for the isothermal case (i.e., $P=\rho C_{\rm s}^2$, where $C_{\rm s}$ is the isothermal sound speed). Numerical simulation of the ambipolar diffusion driven core contraction (Fiedler \& Mouschovias 1993) provided $\kappa \approx 0.47$ which is consistent with a $\kappa$ value of 1/2. 
Ciolek \& Mouschovias (1994) obtained relatively smaller values of $\kappa = 0.38 - 0.43$. 
Mouschovias (1991) suggested that the magnetic field in molecular clouds depends on both the density and the velocity dispersion $\sigma_v$ as $|B| \propto \rho^{1/2} \sigma_v$. Basu (2000) showed that there is a good correlation between $B_{\rm los}$ and $\rho \sigma_v$ in observations, providing $B_{\rm los}/\sigma_v \propto \rho^{0.50 \pm 0.12}$. If the velocity dispersion does not depend on the density, this is consistent with the relation of $|B| \propto \rho^{1/2}$. In contrast, recently Li, McKee \& Klein (2015) conducted a large-scale magneto-hydrodynamic (MHD) simulation of isothermal, self-gravitating gas with a slightly magnetically supercritical initial magnetic field. 
A $\kappa$ value of $0.70 \pm 0.06$ was obtained, and the result is consistent with the value obtained by Crutcher et al. (2010) of $\kappa \approx 2/3$ $(B_{\rm tot} \propto \rho^{0.65 \pm 0.05}$). The flat low density region and the high density region following a power law relation ($\kappa = 0.65$) on the $|B|$ vs. $\rho$ diagram are reproduced in their simulation. Furthermore, it was found that the velocity dispersion scales weakly with density as $\sigma_v \propto \rho^{0.14 \pm 0.05}$, which is also consistent with the result of $\kappa \approx 2/3$. Theoretical studies have revealed a variety of $\kappa$ values ranging from $\kappa \approx 1/2$ to $\kappa \approx 2/3$. Further theoretical studies are desirable in this field. 
\par
%
%
Another critical parameter for magnetic field theories is the ratio of the mass $M$ in the flux tube to the magnitudes of magnetic flux $\Phi$, which is often expressed as the observational parameter normalized by the theoretical critical value, $\lambda = (M/\Phi)_{\rm obs}/(M/\Phi)_{\rm critical}$. Since the magnetic support and the gravity have same radial dependence, the collapse of dense cores cannot be stopped by magnetic fields once gravity overcomes magnetic fields. Theoretical determination of the critical value is thus important. The critical value suggested by theory can be written as $(M/\Phi)_{\rm critical} = c_{\Phi}/\sqrt{G}$, and Mouschovias \& Spitzer (1976) found $c_{\Phi} \approx 0.126$ for disks with support along magnetic field lines. Tomisaka et al. (1988) found a consistent value based on extensive numerical calculations as $c_{\Phi} \approx 0.12$. Nakano \& Nakamura (1978) derived $c_{\Phi}=1/2\pi$ with a linear perturbation analysis for the magnetized isothermal gaseous disk. 
Note that the mass-to-flux ratio depends on cloud geometries, and $(M/\Phi)_{\rm critical} = [3 \pi \sqrt{G/5}]^{-1}$ can be obtained for a uniform sphere under virial equilibrium between gravity and the magnetic field, $3GM^2 / 5R = B^2 R^3 /3$ (Crutcher 2004). Thus, $c_{\Phi} \approx 2/3\pi$ for the spherical case. 
%
Molecular cloud cores in various regions tend to show projected aspect ratios of 2:1 (e.g., Myers et al. 1991; Jijina et al. 1999), and de-projection analyses for revealing the intrinsic shape of dense cores were reported (e.g., Jones et al. 2001: triaxial shape, Tassis et al. 2007: oblate shape with finite thickness). Therefore, in general, observational studies need assumption on the shape of the core when choose and use the theoretical critical value, although the value of $c_{\Phi}=1/2\pi$ (Nakano \& Nakamura 1978) has been widely used. 
\par
Without information of line-of-sight inclination angle of magnetic field direction, $\lambda$ was statistically estimated assuming random orientation of the inclination angle for many target cores. After statistical geometric correction, Crutcher (1999) and Troland \& Crutcher (2008) obtained $\lambda \approx 2$ based on the OH Zeeman observations of dark cloud cores, and the CN Zeeman observations by Falgarone et al. (2008) showed consistent results. Thus, typical dense cores seem to be in a state of slightly magnetically supercritical condition. However, these results have a problem that the statistical analysis eliminates the information of the diversity of the magnetic fields for each core. In order to know $\lambda$ for each core and discuss the magnetic field condition of the core in detail, it is necessary to obtain the information of the magnetic inclination angle $\theta_{\rm inc}$. If $\theta_{\rm inc}$ is known in addition to $\rho$ and $\kappa$, the distribution of $\lambda$ can be obtained from the center of the core to its envelope. As stated by Crutcher (2004), the $\lambda$ value at the cloud envelope provides a crucial test for magnetic support models of star formation. 
\par
In the present study, the $|B|$--$\rho$ relationship was constructed for the starless dense core FeSt 1-457 based on the NIR polarimetric observations of the dichroic polarization of dust toward the background stars. A modified form of the Chandrasekhar--Fermi method, which enables the determination of the value of $\kappa$, was used. With information of the magnetic fields ($\kappa$ and $\theta_{\rm inc}$) and the cloud density distribution, the distribution of mass-to-magnetic flux was obtained, and physical status of FeSt 1-457 was discussed. The mass-to-flux ratio distribution for the case of critical Bonnor--Ebert sphere with $\lambda=2$ was calculated in order to evaluate the behavior of the distribution for typical dense cores. 
\par
FeSt 1-457 is known to be accompanied by an hourglass-shaped magnetic field (Kandori et al. 2017a, hereafter Paper I), and the three dimensional (3D) modeling of the field provides the magnetic field curvature and the line-of-sight inclination angle of the magnetic field direction $\theta_{\rm inc}$ (Kandori et al. 2017b, Paper II). The total magnetic field strength of the core is $33.7 \pm 18.0$ $\mu$G with a ratio of the observed mass-to-magnetic flux to a critical value of $\lambda = 1.41 \pm 0.38$ (magnetically supercritical, Paper II). These analyses seem reliable, because observed NIR polarizations of stars show linear relationship with respect to the dust extinction, indicating that magnetic fields inside FeSt 1-457 is traced by the NIR polarimetry (Kandori et al. 2018, Paper III). The fundamental physical parameters of FeSt 1-457 have been well defined in an internal density structure study based on NIR extinction measurements of the background stars and fitting with the Bonnor--Ebert sphere model (Ebert 1955; Bonnor 1956). The radius, mass, and central density of the core are 18,500 AU (144$''$), $3.55$ $M_{\rm \odot}$, and $3.5 \times 10^{5}$ ${\rm cm}^{-3}$ (Kandori et al. 2005), respectively, at a distance of $130^{+24}_{-58}$ pc (Lombardi et al. 2006). 
\par 
Throughout this paper, the spherical shape was assumed for the core geometry, and $(M/\Phi)_{\rm critical} = 1/2 \pi \sqrt{G}$ (for disk geometry: Nakano \& Nakamura 1978) was used for the theoretical critical mass-to-flux ratio. Though FeSt 1-457 was well fitted using the Bonnor--Ebert sphere model, the elongation in column density structure appears around the core center, which may be the existence of disk-like structure around center. The theoretical critical value for spherical geometry is larger than that for disk geometry, and we thus use the value of $1/2 \pi \sqrt{G}$ as a lower limit of the theoretical critical value. 

\section{Data and Methods}
The NIR polarimetric data for the analysis of the $|B|$--$\rho$ relationship of FeSt 1-457 is taken from Paper I. Observations were conducted using the $JHK$${}_{\rm s}$-simultaneous imaging camera SIRIUS (Nagayama et al. 2003) and its polarimetry mode SIRPOL (Kandori et al. 2006) on the IRSF 1.4-m telescope at the South African Astronomical Observatory (SAAO). SIRPOL can provide deep (18.6 mag in the $H$ band, $5\sigma $ for a one-hour exposure) and wide-field ($7.\hspace{-3pt}'7 \times 7.\hspace{-3pt}'7$ with a scale of 0$.\hspace{-3pt}''$45 ${\rm pixel}^{-1}$) NIR polarimetric data. 
\par
The polarimetry data toward the core is the superposition of the polarizations from both the core itself and the ambient medium which is unrelated to the core. After subtracting the ambient polarization components, 185 stars located within the core radius ($R \le 144''$) in the $H$ band were selected for the polarization analysis (the yellow vectors in Figure 1). 
The most probable configuration of the magnetic field lines pervading the core, estimated using a parabolic function and its rotation, is shown by the solid white lines in Figure 1. The coordinate origin of the parabolic function is fixed to the center of the core measured on the extinction map (R.A.~=~17$^{\rm h}$35$^{\rm m}$47$.\hspace{-3pt}^{\rm s}$5, Decl.~=~$-25^{\circ}$32$'$59$.\hspace{-3pt}''0$, J2000; Kandori et al., 2005). The fitting parameters are $\theta_{\rm mag}=179^{\circ} \pm 11^{\circ}$ and $C = 1.04 (\pm 0.45) \times 10^{-5}$ ${\rm pixel}^{-2}$ ($= 5.14 \times 10^{-5}$ ${\rm arcsec}^{-2}$) for the parabolic function $y = g + gC{x^2}$, where $g$ specifies the magnetic field lines, $\theta_{\rm mag}$ is the position angle of the magnetic field direction (from north to east), and $C$ determines the degree of curvature of the parabolic function (Paper I). 
\par
The parabolic fitting appears to be reasonable because the standard deviation of the residual angles $\theta_{\rm res} = \theta_{\rm obs} - \theta_{\rm fit}$, where $\theta_{\rm fit}$ is the best-fit model position angle, is smaller for the parabolic function ($\delta \theta_{\rm res} = 10.24^{\circ} \pm 0.84^{\circ}$) than for the uniform field case of $16.25^{\circ} \pm 0.70^{\circ}$ (Paper I). 
The intrinsic dispersion can be $\delta \theta_{\rm int} = (\delta \theta_{\rm res}^2 - \delta \theta_{\rm err}^2)^{1/2}$, where $\delta \theta_{\rm err}$ is the standard deviation of the observational error.
\par
In the present study, the radial distribution of the angular difference $\theta_{\rm res}$ is used to derive the magnetic field scaling on density ($|B| \propto \rho^{\kappa}$) toward FeSt 1-457 based on the modified Chandrasekhar--Fermi method and the simple simulations described below. 
\par
Figure 2 shows the simple simulation of the radial distribution of the intrinsic polarization angular difference $\theta_{\rm int}$ (left-hand row of panels) for various values of $\kappa$ in the relationship of $|B| \propto \rho^{\kappa}$ (right-hand row of panels). The horizontal axis in the right-hand row of panels show the line-of-sight mean density, $\rho_{\rm los}$, calculated using the Bonnor--Ebert model with a solution parameter $\xi_{\rm max} = (R/C_{\rm s,eff}) \sqrt{4 \pi G \rho_{\rm c}}= 12.6$, where $R$ is the core radius, $C_{\rm s,eff}$ is the effective sound speed, $G$ is the gravitational constant, and $\rho_{\rm c}$ is the central volume density (Kandori et al. 2005). 
The solid lines in the right-hand row of panels were obtained by calculating average $B_{\rm pos}$ toward each line of sight using the assumed $\kappa$ and known density distribution. Note that the relationship $|B| \propto \rho^{\kappa}$ is not identical to $|B| \propto \rho_{\rm los}^{\kappa}$. Thus, the slope of the relationship on the log $B_{\rm pos}$ -- log $\rho_{\rm los}$ plane is slightly different from the $\kappa$ value in each panel except for the case of $\kappa = 0$. 
%
The $B_{\rm pos}$--$\rho_{\rm los}$ relationships (right-hand row of panels) have the same mean plane-of-sky magnetic field strength of the core (23.8 $\mu$G, Paper I) but have different $\kappa$ indices. The number of data points in each panel in the left-hand row is $N=20,000$, as calculated by generating random numbers following the normal distribution for which the standard deviation is $\delta \theta_{\rm int}$ at each radius. The value of $\delta \theta_{\rm int}$ at each radius was obtained based on the $B_{\rm pos}$--$\rho_{\rm los}$ relationship (right-hand row of panels) and the Chandrasekhar--Fermi formula $\delta \theta_{\rm int} = {C}_{\rm corr} (4 \pi \rho_{\rm los})^{1/2} \sigma_{\rm turb} / {B}_{\rm pos}$ (Chandrasekhar \& Fermi, 1953), where $C_{\rm corr}$ is a correction factor from theory (0.5, Ostriker et al. 2001, see also, Padoan et al. 2001; Heitsch et al. 2001; Heitsch 2005; Matsumoto et al. 2006) and the turbulent velocity dispersion $\sigma_{\rm turb}$ ($0.0573$ km s$^{-1}$, Kandori et al. 2005) was assumed to be constant with respect to the radius. The dot-dashed lines in the left-hand row of panels show $\pm 3 \delta \theta_{\rm int}$.  \par
Figure 2 shows that the radial distributions of $\theta_{\rm int}$ change dramatically from $\kappa = 0$ to $\kappa =1$. Thus, it may not be difficult to determine $\kappa$ directly from the $\theta_{\rm int}$--$r$ diagram, if observational data points are large and accurate. For example, dividing the $\theta_{\rm int}$ data into bins along the radius $r$ and measuring the dispersion $\delta \theta_{\rm int}$ in each bin can produce the radial distribution of $B_{\rm pos} (= {C}_{\rm corr} (4 \pi \rho_{\rm los})^{1/2} \sigma_{\rm turb} / \delta \theta_{\rm int})$ to determine $\kappa$ on the $B_{\rm pos}$--$\rho_{\rm los}$ plane. However, this is not appropriate when the number of data points is limited. The selection of bin size significantly affects the result. Thus, it is important to develop a robust and practical method by which to measure $\kappa$ using a relatively small amount of data. 
\par
Figure 3 shows the results of the simulation for measuring $\kappa$ based on the radial distributions of $\theta_{\rm int}$. The left-hand row of panels and the white solid lines in the right-hand row of panels are the same as those in Figure 2. The radial distribution of the polarization residual angle $\theta_{\rm int}$ (left-hand row of panels) was used to calculate $B_{\rm pos,idv} = {C}_{\rm corr} (4 \pi \rho_{\rm los})^{1/2} \sigma_{\rm turb} / |\theta_{\rm int}|$. In the equation, $|\theta_{\rm int}|$ was used instead of $\delta \theta_{\rm int}$ as in the original Chandrasekhar--Fermi formula, which causes the corresponding dispersion in the derived $B_{\rm pos,idv}$ values on the $B_{\rm pos}$--$\rho_{\rm los}$ plane. The number density distribution of $B_{\rm pos,idv}$ is shown as color images in the right-hand row of panels, in which the distribution of $B_{\rm pos,idv}$ appears flat with respect to $\rho_{\rm los}$ for $\kappa = 0$, and the distributions become steeper for larger $\kappa$. 
%
%
\par
The least squares fitting of the $B_{\rm pos,idv}$ vs. $\rho_{\rm los}$ data with the $|B| \propto \rho^{\kappa}$ relationship (white dashed lines in the right-hand row of panels) provides the index $\kappa$ in the relationship. 
\par
The fitted results (white dashed line) show the same shape, i.e., $\kappa$, but have an offset from the original relationship (white solid line). This is a result of using $|\theta_{\rm int}|$ instead of $\delta \theta_{\rm int}$. The $|\theta_{\rm int}|$ values close to zero cause the large values in $B_{\rm pos,idv}$, so that the resulting $B_{\rm pos,idv}$--$\rho_{\rm los}$ relationship has an upward offset. The existence of the offset is not a problem. The offset can be estimated and removed because both the value of $\kappa$ and the mean magnetic field strength are known. 
\par
The accuracy of the $\kappa$ value depends on the number of stars. In case of $N=20,000$, resulting $\kappa$ values are identical to the original $\kappa$ within $\delta \kappa = 0.01$. For the realistic case of $N=185$, the $\delta \kappa$ increases to $\delta \kappa = 0.10$. 
The method described here provides a robust and practical method by which to determine the $\kappa$ index by a least squares fitting. 

\section{Results and Discussion}
\subsection{Magnetic Field Scaling on Density}
Figure 4 shows the radial distribution of $\theta_{\rm res}$ $(= \theta_{\rm obs} - \theta_{\rm fit})$ for FeSt 1-457. The number of data points is not large ($N=185$) and some outliers exist in the diagram. Thus, it is not easy to determine the $\kappa$ index directly from this figure through fitting. In the present study, the method described in the previous section was used to estimate $\kappa$. 
\par
First, each data point of $\theta_{\rm res}$ was corrected with observational error $\theta_{\rm err}$ in order to derive the estimate of the absolute intrinsic angular difference, $|\theta_{\rm int}| = (|\theta_{\rm res}^2 - \theta_{\rm err}^2|)^{1/2}$. 
Note that the conclusion on the best-fit $\kappa$ value does not change if we mask the data of $\theta_{\rm res}^2 - \theta_{\rm err}^2 < 0$. 
Then, $B_{\rm pos,idv} = {C}_{\rm corr} (4 \pi \rho_{\rm los})^{1/2} \sigma_{\rm turb} / |\theta_{\rm int}|$ was calculated for individual star as a point-to-point application of the Chandrasekhar--Fermi formula. The data points with $|\theta_{\rm int}|$ values close to zero ($|\theta_{\rm int}| < 0.01$) were removed, because such data points produce extremely large $B_{\rm pos,idv}$ values. Note that the number of such data points is small (a few), and this does not change our conclusion. The value of $\sigma_{\rm turb}$ was set to $0.0573$ km s$^{-1}$ measured in the N$_2$H$^+$ ($J=1-0$) line using the Nobeyama 45m radio telescope (Kandori et al. 2005, see also Aguti et al. 2007). The $\sigma_{\rm turb}$ was confirmed to be constant in $r \simlt 70''$ in the core, and thus the relationship was assumed to be flat toward the boundary of the core. 
\par
The obtained $B_{\rm pos,idv}$ vs. $\rho_{\rm los}$ diagram is shown in Figure 5. The solid line shows the least squares fitting of the data points with the $|B| \propto \rho^{\kappa}$ relationship, resulting in $\kappa = 0.78$. This is the first $\kappa$ value estimated toward a single starless core. 
As described in Section 2, the fitting result of $\kappa$ with $N=185$ samples diverges from the original $\kappa$ with uncertainties of $\delta \kappa = 0.10$. 
We evaluated the uncertainty of $\kappa$ using the bootstrap method. A random number following the normal distribution with the same width as the observational error was added for each star, and we performed a least squares fitting. This process was repeated 1,000 times in order to obtain the dispersion of the resulting $\kappa$ values. A value of 0.084 was obtained for the uncertainty of $\kappa$. Considering the estimates of uncertainties, we chose the value of 0.10 for the uncertainty of $\kappa$. 
\par
The obtained value of $\kappa = 0.78 \pm 0.10$ indicates that the case of $\kappa = 2/3$ (isotropic contraction) is preferable for FeSt 1-457, and the difference of the observed value from $\kappa = 1/2$ is 2.8 sigma. A similar value is obtained toward the filamentary cloud IRDC G28.23-00.19 ($\kappa = 0.73 \pm 0.06$, Hoq et al. 2017). These studies support the conclusion by Crutcher et al. (2010). The relatively large $\kappa$ value indicates that the magnetic field in FeSt 1-457 is not very strong. This is consistent with the (slightly) magnetically supercritical feature ($\lambda = 1.41$) of the core. The magnetic field in FeSt 1-457 can be strong enough to control the contraction of the core, because the magnetic field direction of the core ($\theta_{\rm mag}=179^{\circ}$) is perpendicular to the elongation axis of the core ($\theta_{\rm elon} \approx 90^{\circ}$) as shown in Paper I. Observations of ordered magnetic field lines in Figure 1 also support this conclusion. 
\par
The above conditions are consistent with the theoretical MHD simulation results by Li, McKee, \& Klein (2015). They presented two simulation results, one with a slightly magnetically supercritical initial mean field ($\lambda = 1.62$) that is comparable to the parameter of FeSt 1-457, and the other with a very supercritical field ($\lambda = 16.2$). In the former model, well-ordered magnetic field lines appeared in the simulation box and the relatively large value of $\kappa = 0.70 \pm 0.06$ was obtained, which is consistent with the results obtained in the present study. In the latter very weak field model, the magnetic field lines are highly tangled by the turbulent motions, which does not match observations. 
\par
On the basis of the known slope and mean field strength, the plane-of-sky magnetic field strength at the center and boundary of FeSt 1-457 are 93 $\mu$G and 12 $\mu$G, respectively. If we apply the line-of-sight inclination angle of the magnetic field direction ($45^{\circ}$) estimated in Paper II, the total magnetic field strength at the center and boundary of the core are 132 $\mu$G and 17 $\mu$G, respectively. The boundary value of 17 $\mu$G can be used as the estimation of the magnetic field strength in the diffuse inter-clump medium surrounding the core. 
\par
The global plane-of-sky magnetic field strength of the \lq \lq Pipe Bowl'' region, $\approx 2^{\circ}$ around the FeSt 1-457 core, was determined to be 65 $\mu$G (Alves et al. 2008). The value is too large compared with our estimation for the core boundary value. However, this is not surprising because the polarization angle dispersion is integrated and smoothed toward the line of sight in their data. These values are consistent, if we consider the number of (polarization) coherent cell to be $N=30$ (Franco et al. 2010) along the line of sight, resulting in $65 / \sqrt{30} = 11.9$ $\mu$G for the magnetic field strength in each cell. The consistency in the magnetic field strength confirms the coherent cell analysis by Franco et al. (2010). In other words, the present method of the Chandrasekhar--Fermi application to a single core can be used to count the number of line-of-sight polarization coherent cells, if the global magnetic field strength is known. The obtained coherent cell numbers can be compared with the results obtained from the other methods (e.g., Myers \& Goodman 1991; Houde et al. 2009). 
\subsection{Distribution of Mass-to-flux Ratio} 
The distribution of mass-to-flux ratio $\lambda = (M/\Phi)_{\rm obs}/(M/\Phi)_{\rm critical}$ inside FeSt 1-457 was evaluated, since we now know the $\kappa$ index in this paper, mean plane-of-sky magnetic field strength (Paper I), magnetic inclination angle toward the line of sight (Paper II), and density and column density distribution for the core (Kandori et al. 2005). The employed critical value of the mass-to-flux ratio suggested by theory is $1/2\pi G^{1/2}$ (Nakano \& Nakamura 1978). 
\par
First, the offset in the $B_{\rm pos}$--$\rho_{\rm los}$ relationship (solid line in Figure 5) was corrected with known $\kappa = 0.78$ and mean plane-of-sky magnetic field strength for the core of 23.8 $\mu$G (Paper I). Second, the obtained $B_{\rm pos}$ was divided by sin(45$^{\circ}$) (Paper II) to convert it to the total magnetic field strength $B_{\rm tot}$ to obtain the $B_{\rm tot}$--$\rho_{\rm los}$ relationship. Third, since we now know $B_{\rm tot}$ and the column density $N$ for the same line of sight (along the flux tube) is also known (Kandori et al. 2005), the mass-to-flux ratio at each core radius can be obtained by $\lambda = (N/B_{\rm tot})/(1/2\pi G^{1/2})$. 
\par
Caution must be paid at this point. The column density $N$ (Kandori et al. 2005) was measured by subtracting the contribution from ambient medium, and $N$ represents the column density solely associated with the core. The $N$ always goes to zero at core edge, whereas $B$ has a finite value there. Thus, this provides $\lambda = 0$ at core edge, regardless of various $\kappa$ indices. In reality, the place outside the core is not a perfect vacuum and filled with diffuse medium. Magnetic flux tube pervading the core edge region includes the mass of diffuse medium surrounding the core. 
To reflect this effect, we set a cylinder around the core with diameter of $2R$ and height of $2R$ and oriented parallel to the flux tube, and assume that the region outside the core but inside the cylinder is filled with diffuse medium with the density equal to the one at core edge (i.e., $\rho_{\rm diffuse}=\rho_{\rm c}/74.5$, where the coefficient is the density contrast of FeSt 1-457). The $\rho_{\rm diffuse}$ serves as upper limit for ambient diffuse medium, because hotter tenuous gas can achieve pressure-equilibrium at the boundary of the core. 
Since we consider diffuse surrounding medium, we re-calculate average line-of-sight magnetic field strength, $B_{\rm tot}$, by including the contribution from $\rho_{\rm diffuse}$ medium. The $\lambda$ value is then $\lambda = (N+N_{\rm diffuse}) / B_{\rm tot} / (1/2\pi G^{1/2})$. The obtained $\lambda$ represents the averaged value toward the line of sight. 
The result is shown in Figure 6. The $\lambda$ distribution was plotted against the normalized radius (solid line) and the dashed line shows the critical state ($\lambda = 1$). 
\par
The obtained $\lambda$ distribution shows $\approx 2$ toward the core center, and the relationship gradually decreases toward outer region, showing $\lambda \approx 0.98$ at the core edge (nearly magnetically critical or slightly magnetically subcritical). 
The result indicates that FeSt 1-457 is magnetically supercritical inside and critical or slightly subcritical outside. A natural interpretation of this result is that the inter-clump medium surrounding the core is also magnetically critical or slightly subcritical. 
\par
Alves et al. (2008) reported $\lambda_{\rm pos} \approx 0.4$ toward the Pipe Bowl region based on the wide field optical polarization observations. The Pipe Nebula dark cloud complex is known to be less active in star formation except for the spatially limited region around B59 (e.g., Forbrich et al. 2009, 2010), which is consistent with the subcritical feature in the Pipe Bowl region around FeSt 1-457. Since H\,{\sc i} clouds are known to be significantly magnetically subcritical (Heiles \& Troland 2005), it is natural for molecular clouds, assembly of diffuse H\,{\sc i} clouds, to have magnetically subcritical or critical subregions. 
\par
On the basis of these results, we speculate that the FeSt 1-457 core was born from nearly magnetically critical or slightly magnetically subcritical diffuse inter-clump medium. This picture reminds us of the ambipolar diffusion regulated star formation (Shu 1977;  Shu, Adams, \& Lizano 1987). In a classical view, introduced magnetic field strength is very strong, and the magnetically supported cloud core can quasi-statically evolve and reach the singular-isothermal-sphere (SIS) to start inside-out collapse (Shu 1977). In case of moderate magnetic field strength, the cloud core can become magnetically supercritical before reaching the SIS state and start collapse (e.g., Ciolek \& Mouschovias 1994; Ciolek \& Basu 2000). 
Note that though our obtained index of $\kappa = 0.78$ does not fit to the case of strong magnetic fields, it may not be inconsistent with the moderate magnetic field case. 
Though FeSt 1-457 is magnetically supercritical in the central region, this does not mean the core to readily collapse. The theoretical critical mass for the core, $M_{\rm cr} \simeq M_{\rm mag}+M_{\rm BE}$ (Mouschovias \& Spitzer 1976; Tomisaka et al. 1988; McKee 1989), is $3.70\pm0.92$ (Paper II), where $M_{\rm mag}$ is the magnetic critical mass and $M_{\rm BE}$ is the Bonnor--Ebert mass, and the observed core mass, $M_{\rm core} = 3.55\pm0.75$ (Kandori et al. 2005), is identical to $M_{\rm cr}$, suggesting nearly critical state. The magnetically supercritical region in FeSt 1-457 can additionally be supported by thermal pressure, and further magnetic and/or turbulent dissipation should be needed to initiate collapse in the core. Since the core is in a nearly critical state, it is most likely that the magnetohydrostatic configuration (e.g., Tomisaka et al. 1988) can be achieved in FeSt 1-457. The modeling of FeSt 1-457 with respect to the internal density and magnetic field structure is planned. 
\par
It is known that the diffusion timescale $t_{\rm AD}$ is about an order of magnitude longer than the free-fall timescale $t_{\rm ff}$ (e.g., McKee \& Ostriker 2007). This is longer than the observational estimates of the lifetime of prestellar cores (e.g., $\sim 2 - 5$ $t_{\rm ff}$, Ward-Thompson et al. 2007). However, $t_{\rm AD}$ can be shortened by the turbulence and shocks which increase the efficiency of ambipolar diffusion (e.g., Fatuzzo \& Adams 2002; Li \& Nakamura 2004; Kudoh \& Basu 2008). Moreover, in a moderately strong magnetic field case, the cloud core should only lose a part of magnetic flux to become supercritical (Ciolek \& Basu 2000). These effects may bring $t_{\rm AD}$ reasonable length in timescale. 
\par
Finally, we evaluated the distribution of the mass-to-flux ratio for the critical Bonnor--Ebert sphere in order to obtain insights of magnetic criticality for typical dense cores. First, the critical Bonnor--Ebert sphere (temperature $T=10$ K, external pressure $P_{\rm ext}=10^4$ K cm$^{-3}$, and $\xi_{\rm max} = 6.5$) was prepared. The $\lambda$ value toward the center was set to two as a typical value. Second, magnetic field strength toward center was calculated from $2 = N_{\rm center} / B_{\rm center} / (1/2\pi G^{1/2})$. Third, the coefficient $A$ for the relationship $|B| = A \rho^{\kappa}$ was determined by the equation $A = B_{\rm center} / (\int^{R}_{0}\rho(r)^{\kappa}dr) / R$. Forth, the line-of-sight density distribution of the core was manipulated as $A \rho^{\kappa}$ and the quantities were averaged to derive the mean line-of-sight $B$ value. Fifth, according to the same manner of FeSt 1-457 analysis, we set a cylinder around the Bonnor--Ebert core with diameter of $2R$ and height of $2R$ and oriented parallel to the flux tube. The region inside the cylinder but outside the core is filled with diffuse medium with the same density at core edge. The $\lambda$ value for the critical Bonnor--Ebert sphere can be derived as $\lambda = (N + N_{\rm diffuse}) / B / (1/2\pi G^{1/2})$.
%
The results are shown in Figure 7 for the case of $\kappa = 0$ (dotted line), 1/2 (solid line), 2/3 (dashed line), and 1 (dot-dashed line). 
\par
It was reported that the density structure of starless dense cores (globules) can be well fitted by the nearly critical Bonnor--Ebert sphere (Kandori et al. 2005). On the basis of Zeeman observations, $\lambda \approx 2$ was statistically obtained (Crutcher 1999; Troland \& Crutcher 2008; Falgarone et al. 2008). Figure 7 indicates that under the typical input parameters for dense cores (i.e., critical Bonnor--Ebert sphere, $\lambda = 2$, and $\kappa = 1/2$--$2/3$), the distribution of mass-to-flux ratio becomes magnetically critical/subcritical at the core edge. This result implies that the dense cores in previous Zeeman observations are supercritical near the center but critical/subcritical at the edge, and the surrounding diffuse medium of them can also be magnetically critical/subcritical. We thus speculate that typical dense cores are embedded in and evolve from magnetically critical/subcritical diffuse surrounding medium. 
\par
Figure 8 shows the relationship between $\lambda_{\rm edge}$ and $\lambda_{\rm center}$ for $\kappa=$0, 1/2, 2/3, and 1, where $\lambda_{\rm edge}$ is the line-of-sight $\lambda$ value at the core boundary, and $\lambda_{\rm center}$ is the value toward the core center. The lines for each $\kappa$ cross the critical state at $\lambda_{\rm center} \approx 6$ for $\kappa=0$, $\lambda_{\rm center} \approx 2.5$ for $\kappa=1/2$, $\lambda_{\rm center} \approx 2$ for $\kappa=2/3$, and $\lambda_{\rm center} = 1$ for $\kappa=1$. If $\lambda_{\rm center}$ is less than two, the core edge is magnetically subcritical in either case of $\kappa = 1/2$ or $\kappa = 2/3$. We discussed the distribution of mass-to-flux ratio based on the assumption that the geometry of the core is sphere. Exploring the case of other geometry (e.g, spheroid, cylinder, sheet) should be needed for future studies. 
\section{Summary and Conclusion} 
In the present study, the magnetic field scaling on density, $|B| \propto \rho^{\kappa}$, was revealed in a single starless core for the first time. The index $\kappa$ was obtained to be $0.78 \pm 0.10$ toward the starless dense core FeSt 1-457 based on the analysis of the radial distribution of the polarization angle dispersion of background stars measured at the near-infrared wavelengths. The result prefers $\kappa = 2/3$ (isotropic contraction), and the difference of the observed value from $\kappa = 1/2$ is 2.8 sigma. 
The relatively large $\kappa$ value indicates that the magnetic field in FeSt 1-457 is not very strong. This is consistent with the slightly magnetically supercritical feature of the core. The magnetic field in FeSt 1-457 can be strong enough to control the contraction of the core, because the magnetic field direction of the core is perpendicular to the elongation axis of the core. Observations of ordered magnetic field lines around the core also support this conclusion. These results are consistent with the recent theoretical MHD simulation calculated under the slightly magnetically supercritical condition. The total magnetic field strengths at the center and boundary of the core are 132 $\mu$G and 17 $\mu$G, respectively. The boundary value can be used as the estimation of the magnetic field strength in the diffuse inter-clump medium surrounding the core. 
%
%
On the basis of $\kappa$ and known density structure, the distribution of the ratio of mass to magnetic flux was evaluated. FeSt 1-457 was found to be magnetically supercritical near the center ($\lambda \approx 2$), whereas nearly critical (slightly subcritical) at the core boundary ($\lambda \approx 0.98$). Thus, the diffuse inter-clump medium surrounding the core can also be nearly magnetically critical. Ambipolar diffusion regulated star formation models for the case of moderate magnetic field strength may explain the physical status of FeSt 1-457. Note that though our obtained index of $\kappa = 0.78$ does not fit to the case of strong magnetic fields, it may not be inconsistent with the moderate magnetic field case. The mass-to-flux ratio distribution for typical dense cores (critical Bonnor--Ebert sphere with central $\lambda=2$ and $\kappa=1/2$--$2/3$) was found to be magnetically critical/subcritical at the core edge, which indicates that typical dense cores are embedded in and evolve from critical/subcritical diffuse surrounding medium. 


\subsection*{Acknowledgement}
We are grateful to the staff of SAAO for their kind help during the observations. We wish to thank Tetsuo Nishino, Chie Nagashima, and Noboru Ebizuka for their support in the development of SIRPOL, its calibration, and its stable operation with the IRSF telescope. The IRSF/SIRPOL project was initiated and supported by Nagoya University, National Astronomical Observatory of Japan, and the University of Tokyo in collaboration with the South African Astronomical Observatory under the financial support of Grants-in-Aid for Scientific Research on Priority Area (A) Nos. 10147207 and 10147214, and Grants-in-Aid Nos. 13573001 and 16340061 of the Ministry of Education, Culture, Sports, Science, and Technology of Japan. RK, MT, NK, KT (Kohji Tomisaka), and MS also acknowledge support by additional Grants-in-Aid Nos. 16077101, 16077204, 16340061, 21740147, 26800111, 16K13791, 15K05032, and 16K05303.

\clearpage 

\begin{figure}[t]  
\begin{center}
 \includegraphics[width=6.5 in]{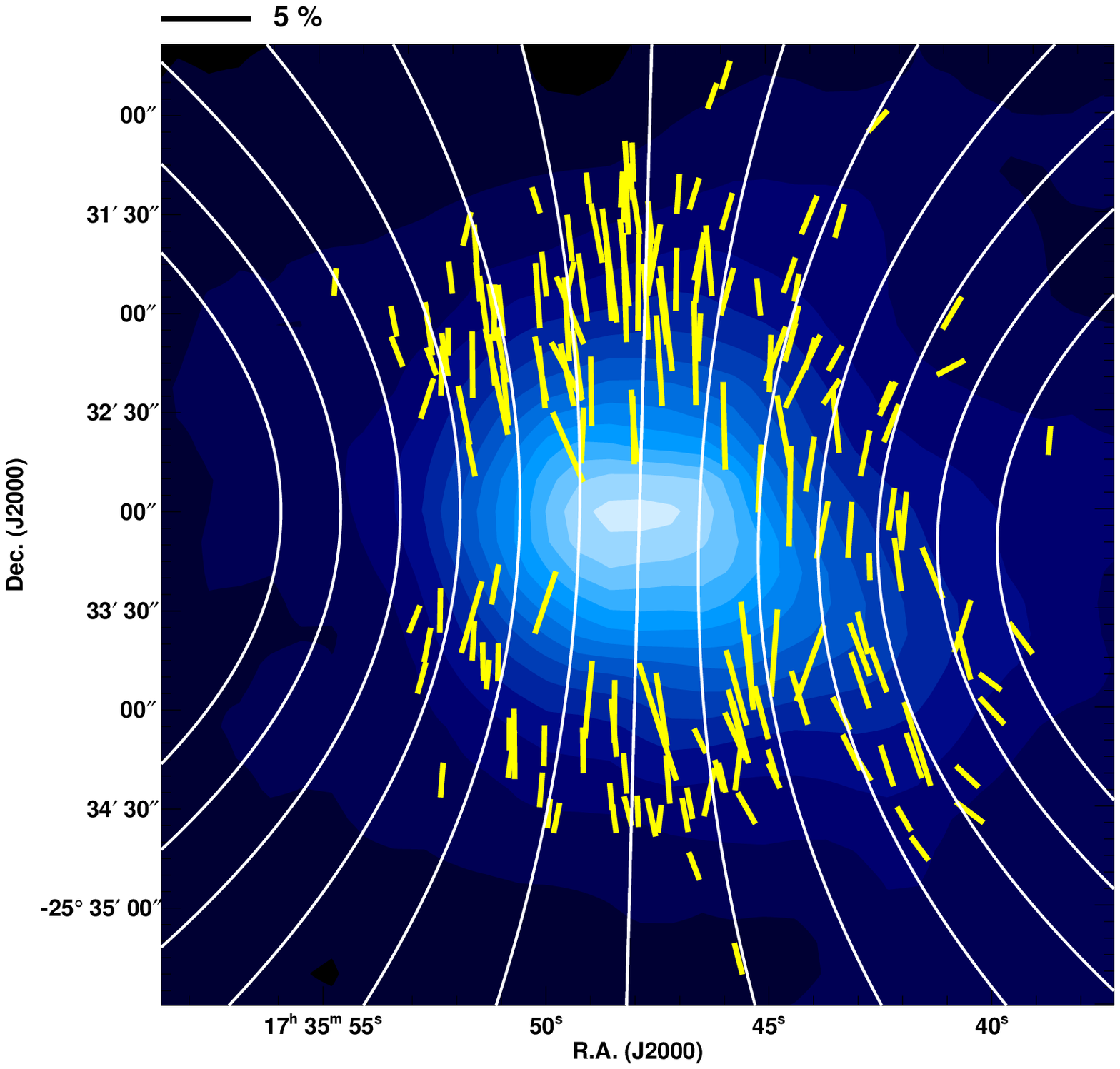}
 \caption{Polarization vectors of FeSt 1-457 after subtracting the ambient polarization component. The field of view is the same as the diameter of the core ($288''$ or 0.19 pc). The white lines indicate the magnetic field direction inferred from the fitting with a parabolic function of, $y = g + gC{x^2}$, where $g$ specifies the magnetic field lines and $C$ determines the degree of curvature in the parabolic function. The scale of the 5$\%$ polarization degree is shown at the top. 
The background image is the $A_V$ distribution taken from Kandori et al. (2005). In the image, grey scale (filled contour) starts from $A_V=0$ mag with a step of 3 mag. The resolution of the $A_V$ measurements is $33''$.}
   \label{fig1}
\end{center}
\end{figure}

\clearpage 

\begin{figure}[t]  
\begin{center}
 \includegraphics[width=6.0 in]{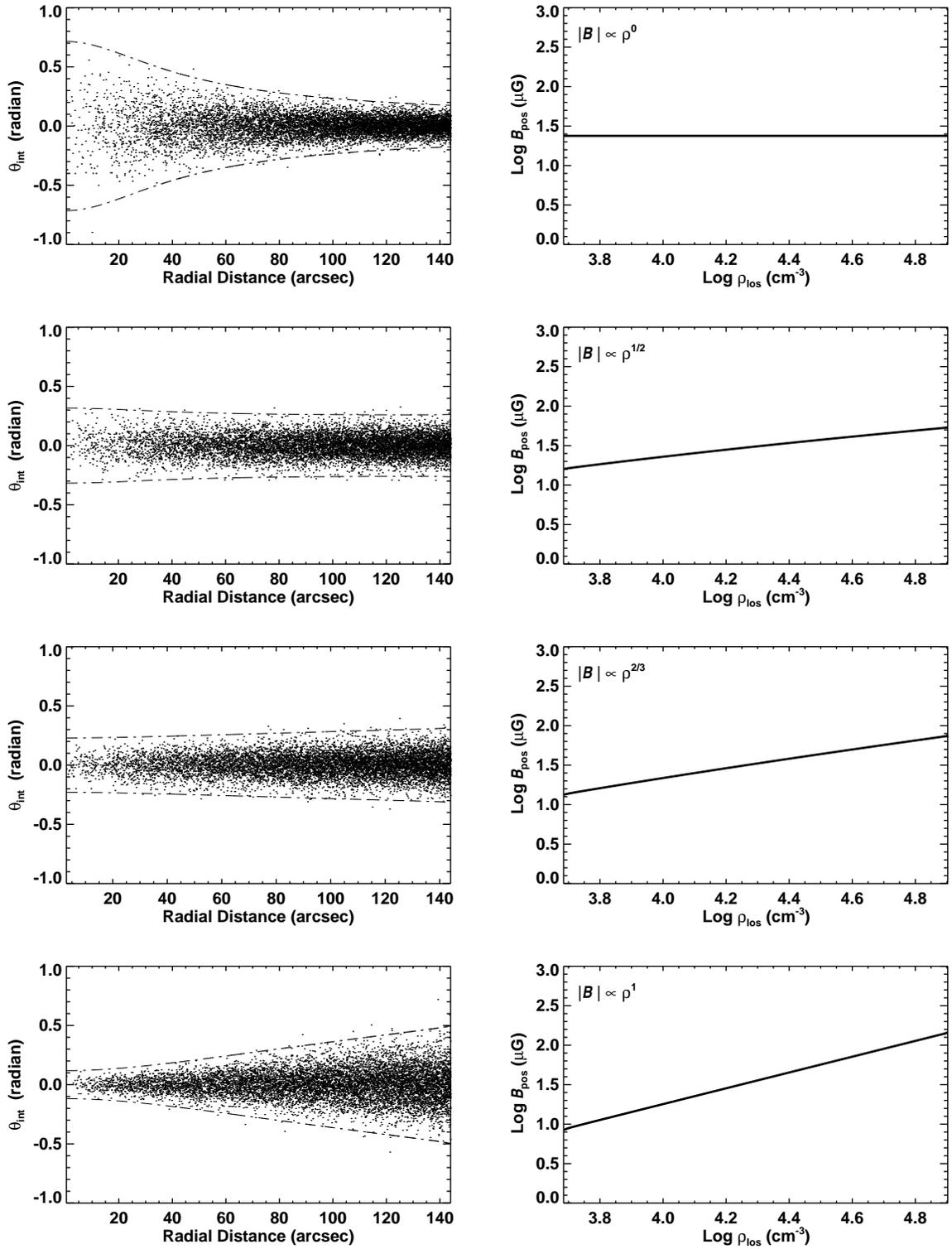}
 \caption{A simulation of the radial distribution of the polarization residual angle $\theta_{\rm int}$ (left-hand row of panels) for various values of $\kappa$ in the relationship of $|B| \propto \rho^{\kappa}$ (right-hand row of panels). The horizontal axis in the right-hand row of panels shows the line-of-sight mean density, $\rho_{\rm los}$, calculated using the Bonnor--Ebert sphere model with a solution parameter of $\xi_{\rm max}=12.6$. 
%
The number of data points in each panel in the left-hand row is $N=20,000$ calculated by generating a random number following a normal distribution with a standard deviation of $\delta \theta_{\rm int}$ at each radius. The value of $\delta \theta_{\rm int}$ at each radius was obtained based on the Chandrasekhar--Fermi formula and the $B_{\rm pos} - \rho_{\rm los}$ relationship in the right-hand row of panels. The dot-dashed lines in the left-hand row of panels show $\pm 3 \delta \theta_{\rm int}$. The horizontal axis in each panel corresponds to the core center to the boundary of $r=144''$.
}
   \label{fig1}
\end{center}
\end{figure}

\clearpage 

\begin{figure}[t]  
\begin{center}
 \includegraphics[width=6.0 in]{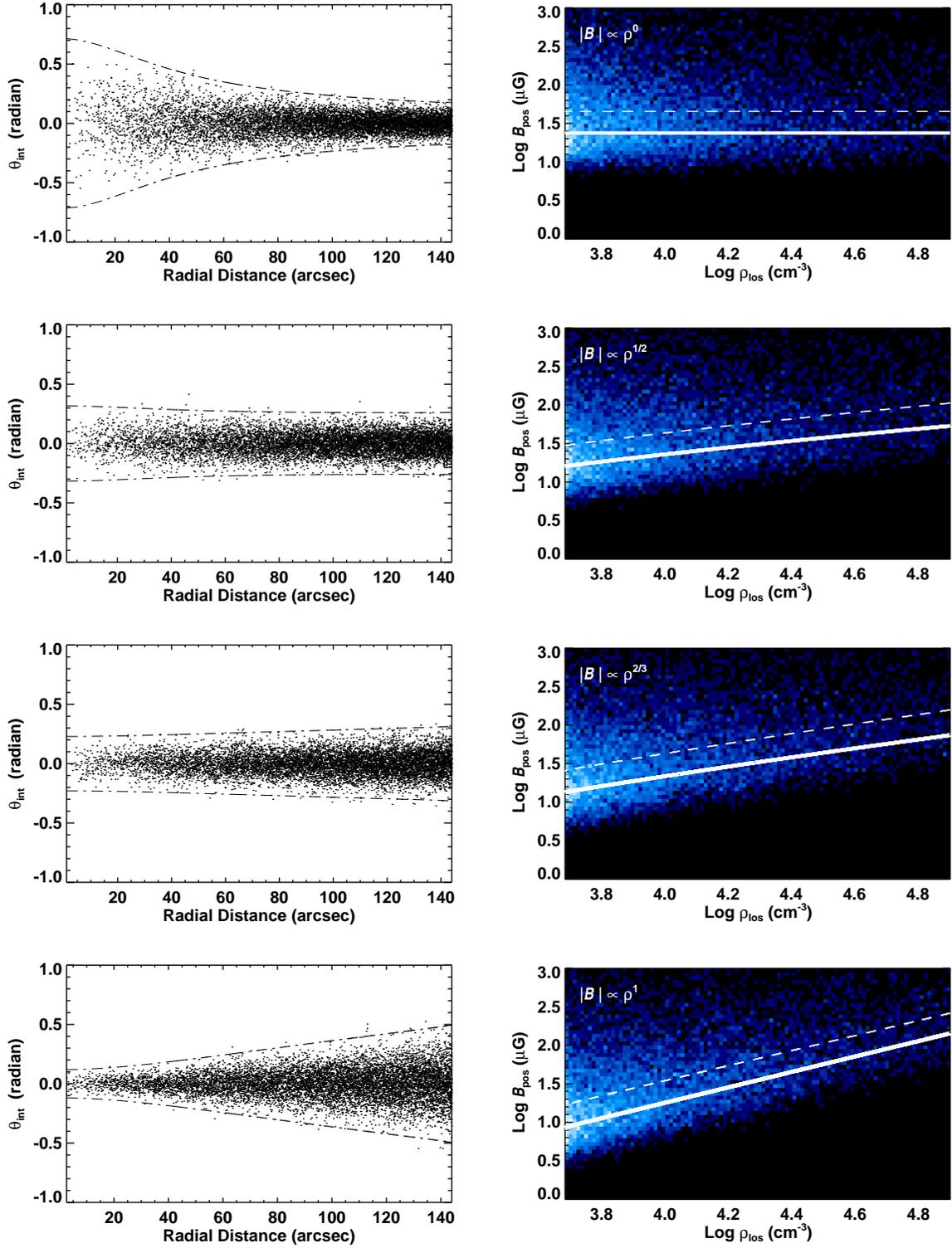}
 \caption{Same as Figure 2, but the relationship of $|B| \propto \rho^{\kappa}$ is indicated by the white solid lines in the right-hand row of panels. The radial distribution of the polarization residual angle $\theta_{\rm int}$ (left-hand row of panels) was used to calculate $B_{\rm pos,idv} = {C}_{\rm corr} (4 \pi \rho_{\rm los})^{1/2} \sigma_{\rm turb} / |\theta_{\rm int}|$. The number density distributions of $B_{\rm pos,idv}$ are shown as color images in the right-hand row of panels. Their least squares fits are indicated by the white dashed lines in the right-hand row of panels, whereas the white solid lines indicate the original relationship. The offset between the two lines can be removed because the value of $\kappa$ and the mean magnetic field strength of the core are known.
}
   \label{fig1}
\end{center}
\end{figure}

\clearpage 

\begin{figure}[t]  
\begin{center}
 \includegraphics[width=6.5 in]{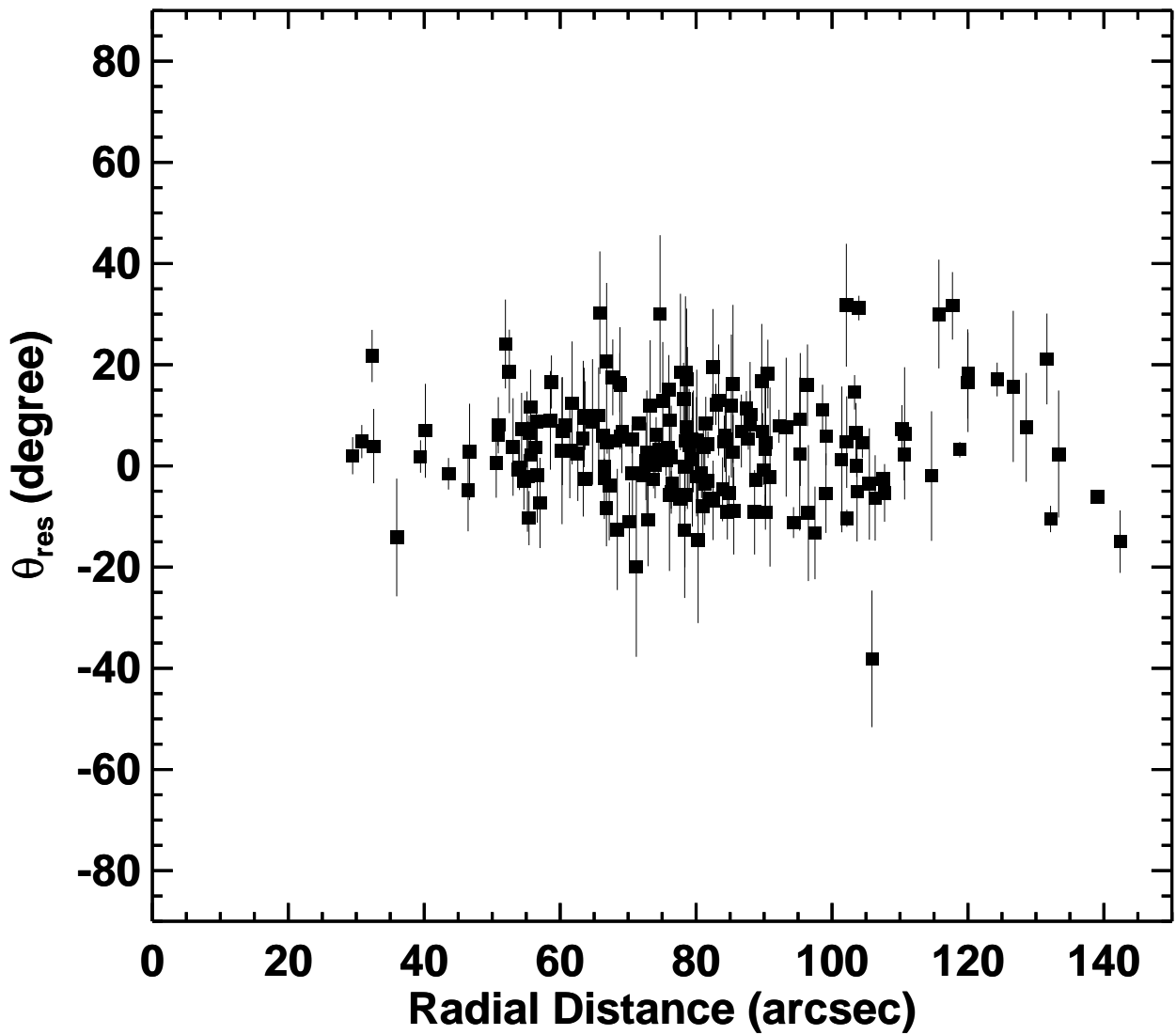}
 \caption{Distribution of the residual polarization angle $\theta_{\rm res}$ ($= \theta_{\rm obs} - \theta_{\rm fit}$) obtained using a parabolic function. The horizontal axis ranges from the core center to the boundary of $r=144''$.}
   \label{fig1}
\end{center}
\end{figure}

\clearpage 

\begin{figure}[t]  
\begin{center}
 \includegraphics[width=6.5 in]{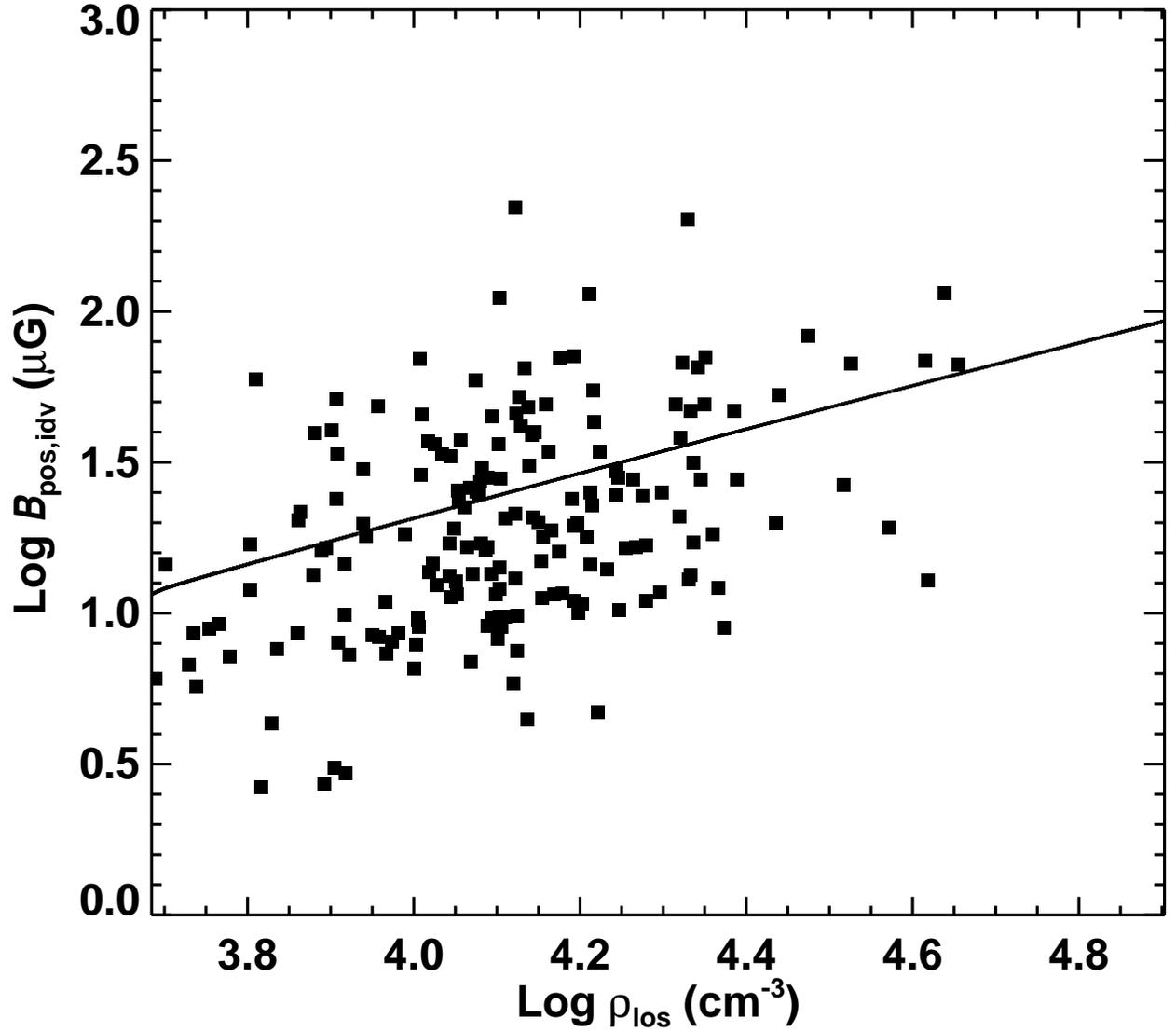}
 \caption{Distribution of $B_{\rm pos,idv}$ calculated from the observed residual polarization angle 
$\theta_{\rm res}$. The solid line shows the least squares fit of the data based on the relationship of $|B| \propto \rho^{\kappa}$.}
   \label{fig1}
\end{center}
\end{figure}	

\clearpage 

\begin{figure}[t]  
\begin{center}
 \includegraphics[width=6.5 in]{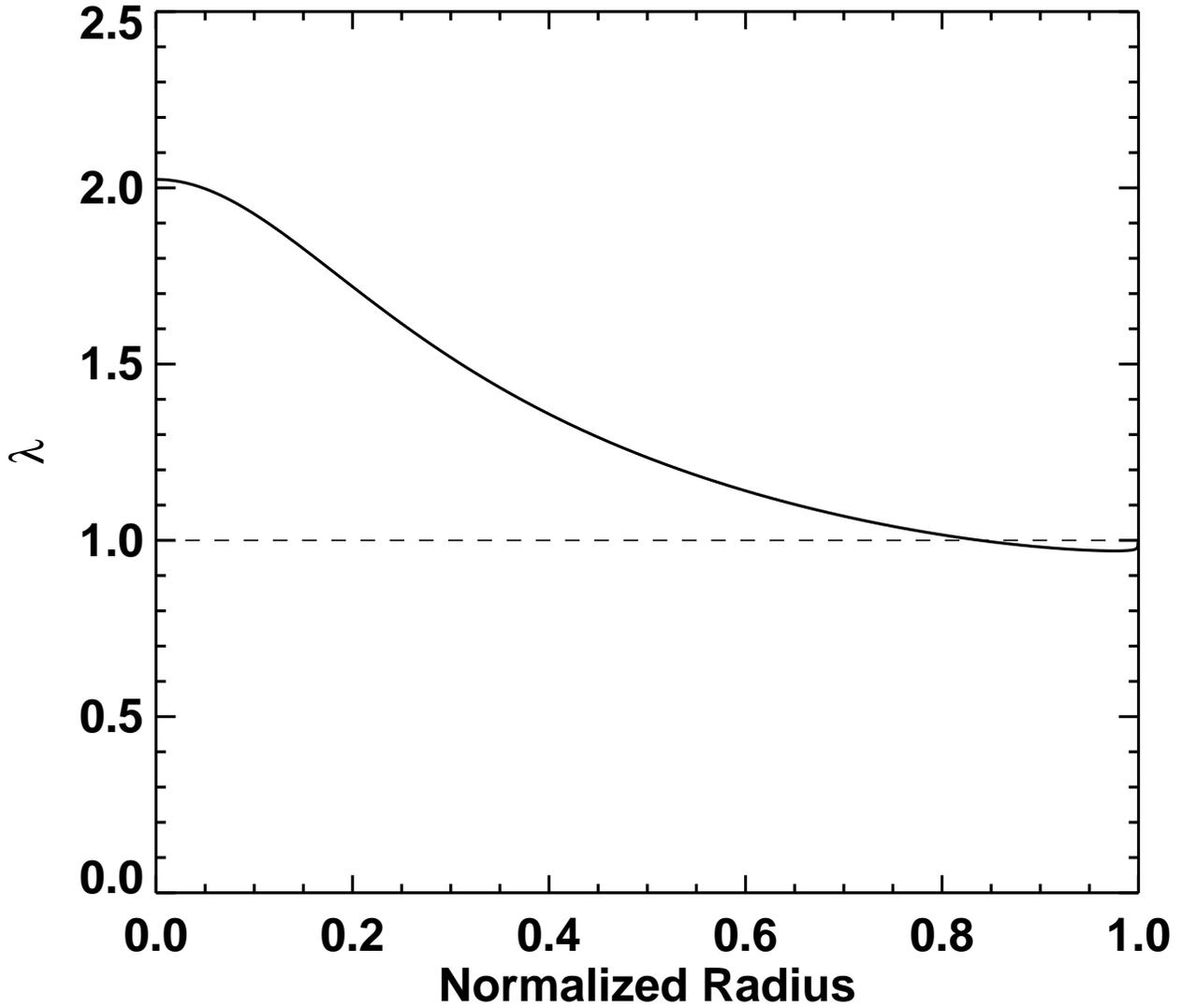}
 \caption{ 
 Distribution of $\lambda$ toward the lines of sight calculated from known density structure of FeSt 1-457 and $\kappa = 0.78$ (solid line). The dashed line shows the critical state ($\lambda = 1$).
 }
   \label{fig1}
\end{center}
\end{figure}	

\clearpage 

\begin{figure}[t]  
\begin{center}
 \includegraphics[width=6.5 in]{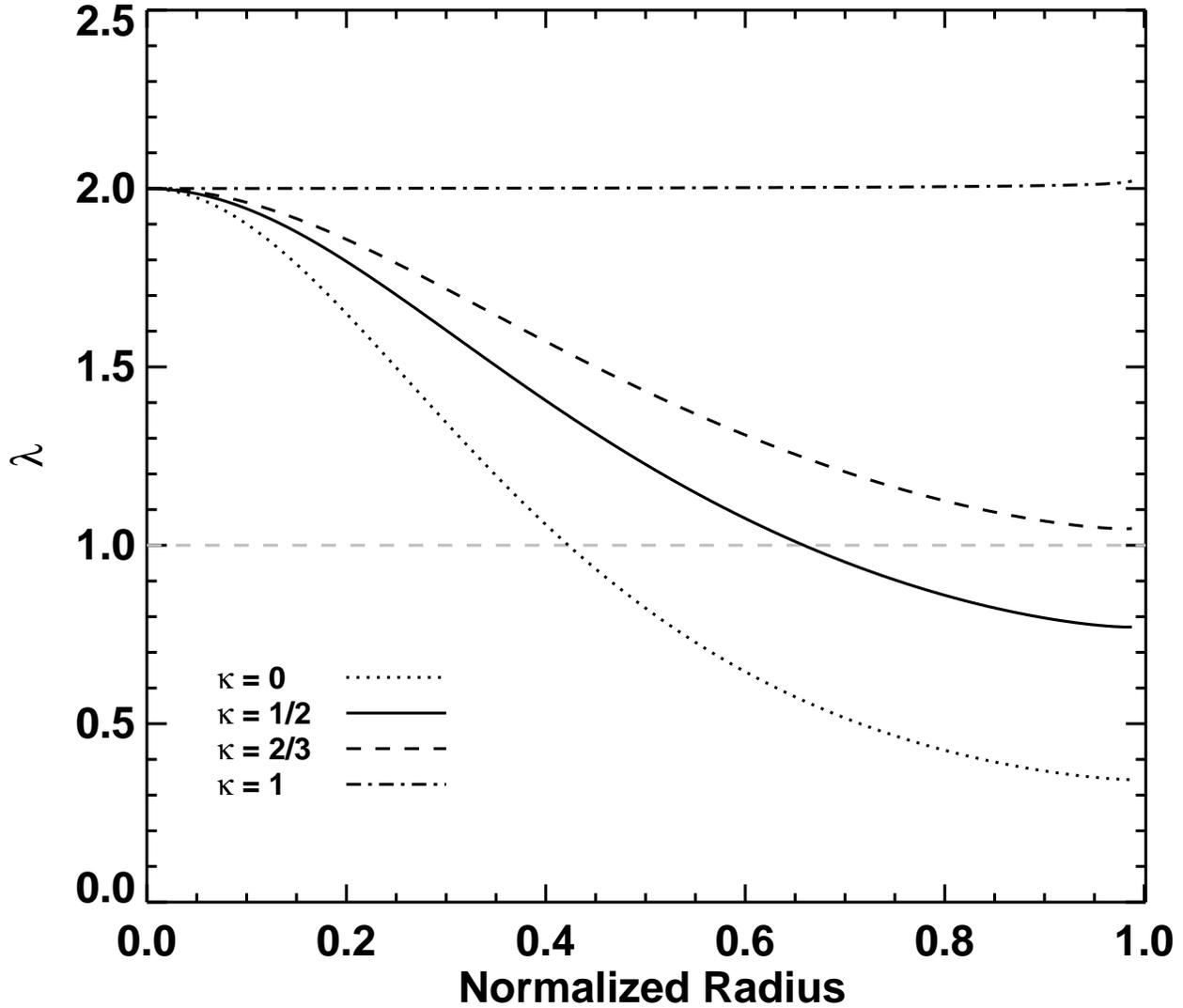}
 \caption{ 
 Distribution of the line-of-sight $\lambda$ values calculated for the case of the critical Bonnor--Ebert sphere. The value of $\lambda$ toward the core center is set to two. The dotted line, solid line, dashed line, and dot-dashed line correspond to $\kappa=0$, 1/2, 2/3, and 1, respectively. The grey dashed line shows the critical state ($\lambda = 1$).
 }
   \label{fig1}
\end{center}
\end{figure}	

\clearpage 

\begin{figure}[t]  
\begin{center}
 \includegraphics[width=6.5 in]{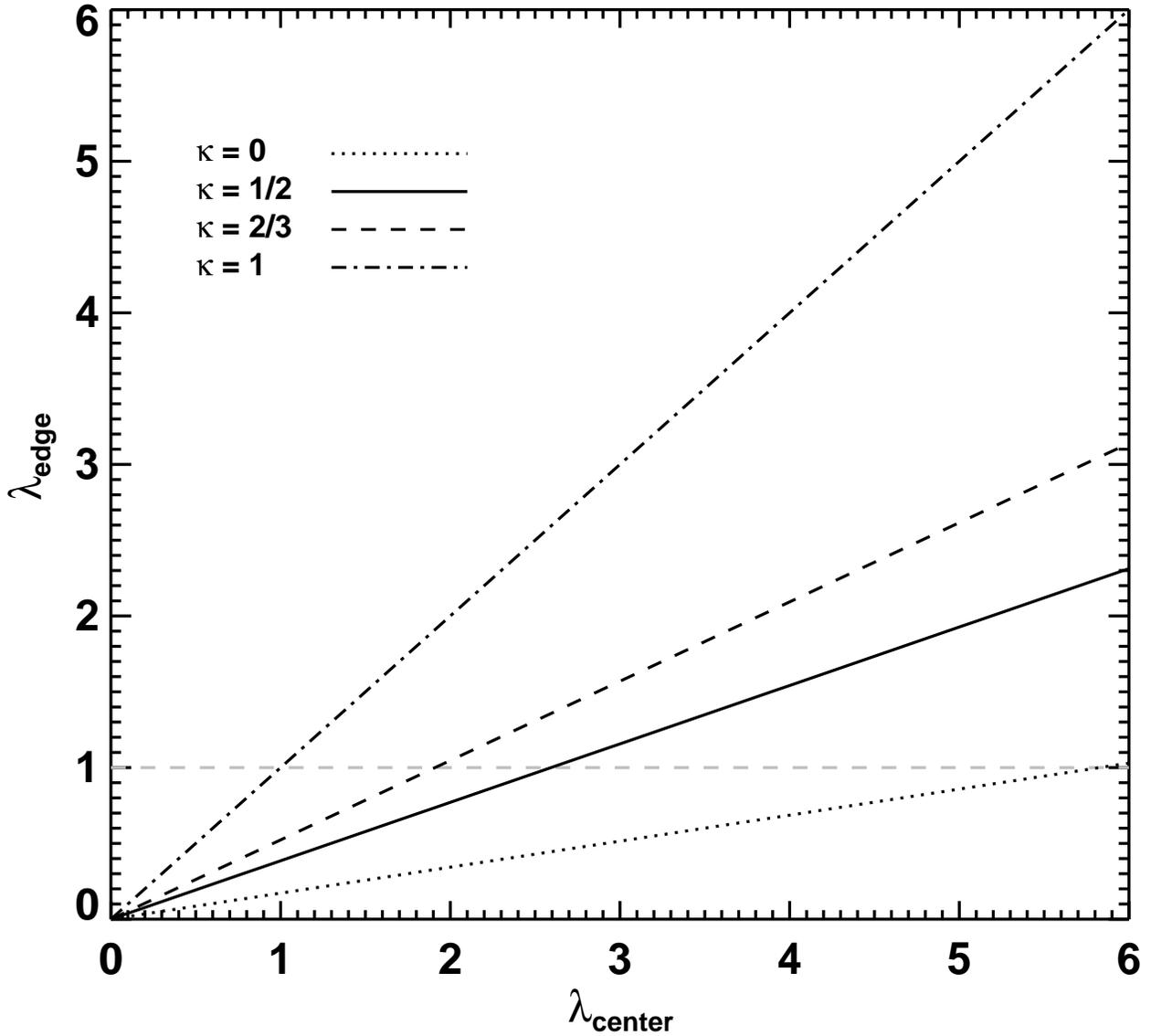}
 \caption{ 
The relationship between $\lambda_{\rm edge}$ and $\lambda_{\rm center}$, where $\lambda_{\rm edge}$ is the line-of-sight $\lambda$ value at core boundary, and $\lambda_{\rm center}$ is the value toward core center. This diagram was made for the case of the critical Bonnor--Ebert sphere. The dotted line, solid line, dashed line, and dot-dashed line correspond to $\kappa=0$, 1/2, 2/3, and 1, respectively. The grey dashed line shows the critical state ($\lambda = 1$).
}
   \label{fig1}
\end{center}
\end{figure}	

\end{document}